# Dislocation density distribution at slip band-grain boundary intersections

Yi Guo[a]*, David M. Collins[b], Edmund Tarleton[c,d], Felix Hofmann[d], Angus J. Wilkinson[c], T. Ben Britton[a]

[a] Department of Materials, Royal School of Mines, Imperial College London, London SW7 2AZ, UK

[b] School of Metallurgy and Materials, University of Birmingham Edgbaston, Birmingham, B15 2TT

[c] Department of Materials, University of Oxford, Parks Road, Oxford OX1 3PH, UK

[d] Department of Engineering Science, University of Oxford, Parks road, Oxford, OX2 7TL, UK

**Abstract**

We study the mechanisms of slip transfer at a grain boundary, in titanium, using Differential Aperture X-ray Laue Micro-diffraction (DAXM). This 3D characterisation tool enables measurement of the full (9-component) Nye lattice curvature tensor and calculation of the density of geometrically necessary dislocations (GNDs). We observe dislocation pile-ups at a grain boundary, as the neighbour grain prohibits easy passage for dislocation transmission. This incompatibility results in local micro-plasticity within the slipping grain, near to where the slip planes intersect the grain boundary, and we observe bands of GNDs lying near the grain boundary. We observe that the distribution of GNDs can be significantly influenced by the formation of grain boundary ledges that serve as secondary dislocation sources. This observation highlights the non-continuum nature of polycrystal deformation and helps us understand the higher order complexity of grain boundary characteristics.

**Graphical Abstract:**

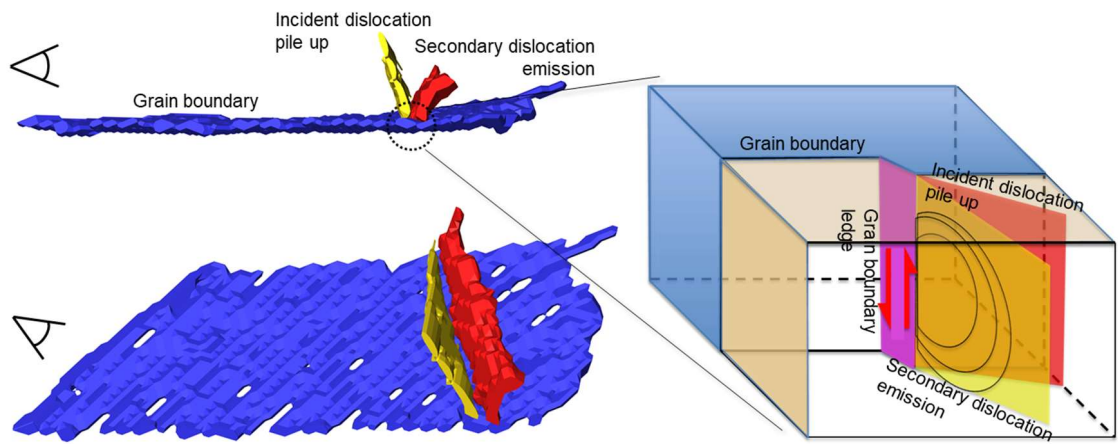

# 1. Introduction

Most structural metals and alloys used are polycrystalline with grain boundaries typically imparting significant strengthening. The field of grain boundary engineering expressly explores how specific boundary types can be introduced during thermomechanical processing, however, experimental evidence of the relative benefit of different boundary types is limited. Plastic deformation of a polycrystalline aggregate typically results in complex heterogeneous deformation patterns, due to the compatibility and equilibrium requirements as well as the heterogenous dislocation sources and the blocking of dislocation slip. A physically based model for intergranular deformation must account not only for the mechanical compatibility across the grain boundary [1], but also the force equilibrium [2,3]. These two conditions are important when we consider the heterogenous deformation fields that occur as a result of slip on individual slip planes, especially where they impinge on grain boundaries.

Because of the localised slip, deformation near a grain boundary is heterogenous and the equilibrium and compatibility conditions can require the generation of dislocations near the interface. These must accommodate the strain incompatibility and overcome variations in the elastic properties across the interface. In micromechanics, the storage of dislocations has been considered by Ashby [4] who describes dislocation densities as either statistically stored (SSD) or geometrically necessary (GND). While the GND density is considered as a means to accommodate orientation gradients and to maintain lattice continuity, the SSD density represents dislocation dipoles or multipoles randomly trapped during plastic deformation, which, at the length-scale of inquiry, do not contribute to a net lattice curvature. This concept was explained in further detail by Arsenlis and Parks [5]. Formally, if we could look with fine enough resolution at each individual dislocation, they would all be considered geometrically necessary due to the associated closure failure of the Burgers circuit and therefore the classification of dislocation densities into GND and SSD contributions is dependent on the length scale of the measurement.

To assist in the assessment of the GND density, Nye's dislocation tensor [6] combines dislocation content on various slip systems into one tensor

(1) $\boldsymbol{\alpha} = \sum_k \rho^k \boldsymbol{b}^k \otimes \boldsymbol{l}^k$

where $\rho^k, \boldsymbol{b}^k, \boldsymbol{l}^k$ are the GND density, Burgers vector, and unit line direction of the $k^{th}$ dislocation type respectively. Kröner [7] and Bilby et al. [8] demonstrated that the Nye's tensor can be linked to the elastic distortions of the matrix through:

(2) $\boldsymbol{\alpha} = \nabla \times \boldsymbol{\beta}$

Where the $\nabla \times$ is a Curl operator. $\beta$ is the elastic displacement gradient of the matrix, which can be measured by X-ray diffraction, electron backscatter diffraction (EBSD), or other techniques.

Prior research has attempted to investigate the accumulation of GND density in several crystal systems subjected to various deformation conditions. They aimed to rationalize aspects of fundamental material behaviour including; the Hall-Petch effect [9,10], the strain hardening [11], and the indentation size effect [12–14]. One common way of estimating the GND density is to extract the local orientation information using EBSD methods [15–19]. Such measurements can be improved by increasing the angular precision of the data with high angular resolution EBSD (HR-EBSD) [20–26].

For the calculation of the Nye tensor, the elastic displacement gradient must be measured. This requires a measurement of the spatial derivative of the elastic lattice distortion. If one assumes the elastic strain gradient is negligible, surface measurements (e.g. EBSD/HR-EBSD) only provide five terms of the Nye tensor and one difference ($a_{11} - a_{22}$) [17]. It is possible to investigate the lattice curvatures in the third dimension through focused ion beam serial sectioning, as has been demonstrated by some pioneering studies [13,27]. However, both the angular and the translational alignment errors between slices may introduce additional noise.

Full three-dimensional characterisation of the Nye tensor can also be performed non-destructively using high-energy X-ray diffraction. Some state-of-the-art methods for this purpose include the Differential Aperture X-ray Laue Micro-diffraction (DAXM) [28] and the emerging Dark-Field X-ray microscopy [29]. Both of the techniques have submicron spatial resolution and the angular resolution is on the order of 0.01°.

In this paper, we use DAXM to investigate in 3D the local dislocation structure generated by the interaction between slip bands and a grain boundary in commercially pure titanium. We first introduce the process to retrieve Nye's lattice curvature tensor from the measured lattice rotations in 2D and 3D. This is then compared to the resultant dislocation density obtained directly from the entrywise 1-norm of the curvature tensor without linking to the dislocation tensor [15]. Based on the observed distribution of dislocation densities, we investigate the mechanism of the slip band-grain boundary interactions under the framework of the grain boundary ledge theory put forward by Li [30] and Hirth [31] as well as referring to the general requirement of elastic and plastic compatibility at a grain boundary. This investigation is fundamental to the understanding and simulation of heterogeneous deformation at a polycrystalline interface [32–34].

## 2. Experiments

The material used in this experiment was Grade I commercial purity titanium. Details of the sample composition and preparations can be found elsewhere [25]. This sample was investigated using HR-EBSD (after deformation in tension to 0.01 strain) at various locations where a slip band intersects a grain boundary. A blocked slip band that had shown evidence of distinct GND density concentration [25] was selected as a suitable candidate for further DAXM investigation.

The DAXM experiment was performed at beam line 34-ID-E at the Advanced Photon Source (APS). A detailed overview of the technique, analysis method and some applications can be found in references [28,35–37], and the technical details of the current experiment can be found in reference [35]. A brief overview is given here:

The sample was positioned in the X-ray beam and tilted to ~45° for positioning the differential aperture (a Pt wire). A focused white beam was directed towards the selected region and was scanned on a regular grid with 1 μm spacing. At each scanning node, the differential aperture swept across the sample surface and occluded the diffraction spots. The voxel-based diffraction patterns were reconstructed by ray-tracing using the LaueGo software (performed on the cluster at the APS). Reconstructions were performed to a depth of 190 μm into the sample with 1 μm depth spacing. The total volume of investigation is of the size 23 μm x 31 μm x 190 μm, as is shown in Fig. 1, with 1 μm$^3$ voxel size. The lengths of the <c> and <a> axis for Ti were calibrated using the monochromatic beam giving a measured c/a ratio of 1.588. In addition to extracting the lattice parameters, a reliability factor based upon the number of spots used in the indexing was obtained. This was used to filter out erroneous 'ghost points' from above the sample surface, observed due to the nature of the back-subtraction algorithm from the DAXM analysis procedure. The indexed dataset was analysed with Igor Pro (www.wavemetrics.com) where the indexed lattice parameters were compared to a user defined reference point in each grain, and hence relative lattice rotations (***R***) were obtained (including reduction to the symmetry-reduced rotations). The white beam measurements enabled only the deviatoric deformation to be obtained (i.e. hydrostatic strains are not measured). The lattice rotations and the strains were exported into MATLAB (www.mathworks.com) for further postprocessing and the 3D images were processed in Avizo (www.thermofisher.com).

## 3. Results

The illuminated volume of the sample is shown in Fig. 1-a where the traces of the grain boundary and the slip bands are also identified. Fig. 1-b shows a slice view of the grain boundary plane, identifying the lines of intersections between the slip planes and grain boundary as evident from the high intensity dislocation bands. In total, 6 dislocation bands are seen in Fig. 1-b, corresponding to 6 slip plane and grain boundary interactions. From Fig. 1, it is noted that all 6 slip planes possess the same slip variant (i.e. prismatic slip with the same plane and Burgers vector indices), however, evidently possess different levels of plasticity at the grain boundary.

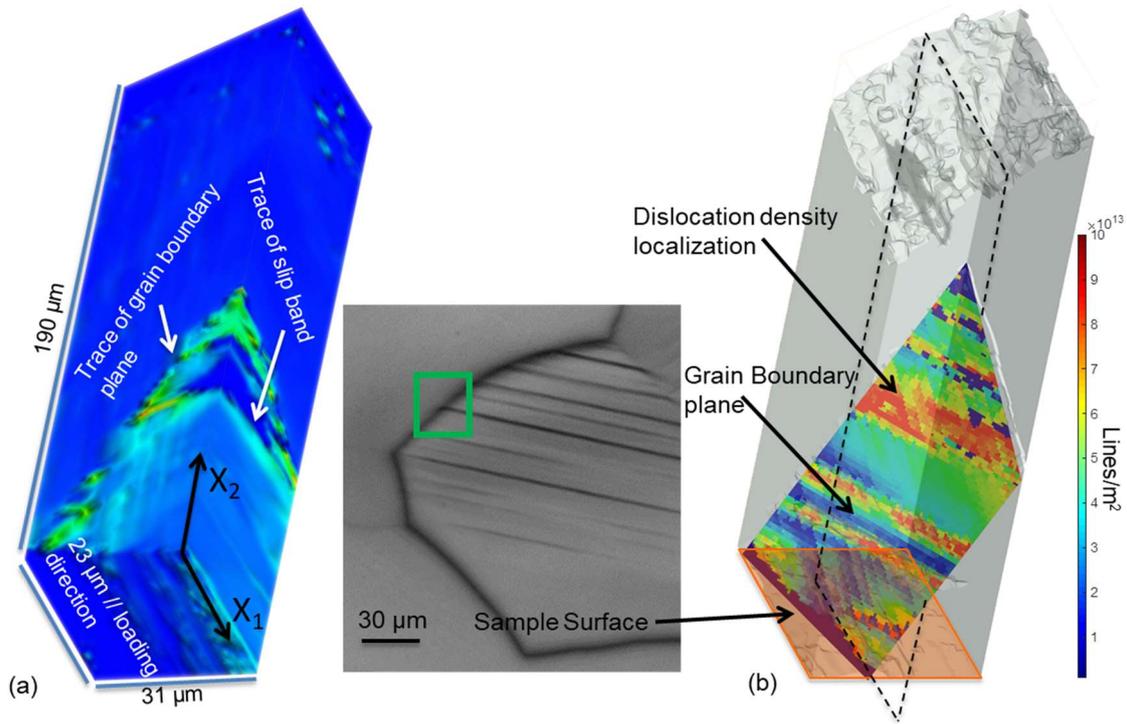

*Figure 1: a) Visualization of the probed volume, color coded with respect to the GND density, with an inset optical micrograph indicating the sampling location. The grain boundary trace and the observed slip band is indicated by white arrows; b) a virtual slice taken of the grain boundary plane (the rest made transparent), high intensity bands, representing localization of GND densities, indicating the location where slip planes intersect the grain boundary. The dashed box indicates the location of the slice taken for the rest of the analysis. The viewing angle (i.e. camera position) in Figure 1 is chosen to best view the grain boundary plane and the slip traces, for a better visualization of the shape of the 3D volume, the reader is referred to the animation in the Appendix A2'.*

## 3.1 GND density calculation

The method for GND density calculation follows that developed for HR-EBSD [20,38,39], i.e. solving the Nye-Kröner-Bilby (NKB) equation (equations 1 and 2) but extended to incorporate the gradients in the third dimension, therefore accounting for all nine components of the elastic displacement gradient tensor [6,40]:

(3) $\alpha_{ij} = \alpha_{ij}^{\omega} + \alpha_{ij}^{\varepsilon} =$

$$\begin{pmatrix} \frac{\partial \omega_{12}}{\partial x_3} - \frac{\partial \omega_{13}}{\partial x_2} & \frac{\partial \omega_{13}}{\partial x_1} & \frac{\partial \omega_{21}}{\partial x_1} \\ \frac{\partial \omega_{32}}{\partial x_2} & \frac{\partial \omega_{23}}{\partial x_1} - \frac{\partial \omega_{21}}{\partial x_3} & \frac{\partial \omega_{21}}{\partial x_2} \\ \frac{\partial \omega_{32}}{\partial x_3} & \frac{\partial \omega_{13}}{\partial x_3} & \frac{\partial \omega_{31}}{\partial x_2} - \frac{\partial \omega_{32}}{\partial x_1} \end{pmatrix} + \begin{pmatrix} \frac{\partial \varepsilon_{12}}{\partial x_3} - \frac{\partial \varepsilon_{13}}{\partial x_2} & \frac{\partial \varepsilon_{13}}{\partial x_1} - \frac{\partial \varepsilon_{11}}{\partial x_3} & \frac{\partial \varepsilon_{11}}{\partial x_2} - \frac{\partial \varepsilon_{12}}{\partial x_1} \\ \frac{\partial \varepsilon_{22}}{\partial x_3} - \frac{\partial \varepsilon_{23}}{\partial x_2} & \frac{\partial \varepsilon_{23}}{\partial x_1} - \frac{\partial \varepsilon_{21}}{\partial x_3} & \frac{\partial \varepsilon_{21}}{\partial x_2} - \frac{\partial \varepsilon_{22}}{\partial x_1} \\ \frac{\partial \varepsilon_{32}}{\partial x_3} - \frac{\partial \varepsilon_{33}}{\partial x_2} & \frac{\partial \varepsilon_{33}}{\partial x_1} - \frac{\partial \varepsilon_{31}}{\partial x_3} & \frac{\partial \varepsilon_{31}}{\partial x_2} - \frac{\partial \varepsilon_{32}}{\partial x_1} \end{pmatrix}$$

where $\omega_{ij}$ is lattice rotation and $\varepsilon_{ij}$ elastic strain. Equation (3) is used to solve Nye's dislocation tensor in equation (1), for which purpose the lattice rotations and the elastic strains need to be measured.

The rotations and the strains are obtained by post processing the experiment results as stated above. A local cubic kernel containing 27 voxels was extracted from the data volume centred at each interrogation point. The data in the kernel was filtered to eliminate those voxels from outside the material volume or from a different grain. After filtering, a 'useful' kernel should contain, in addition to the central voxel, at least one voxel in each of the principal reference directions to capture the lattice rotation gradient associated with that direction. Then a disorientation matrix [22] was calculated between each voxel in the kernel and the centre voxel of the kernel:

(4) $\boldsymbol{\omega_{ij}} = \boldsymbol{R}_{centre}^T \boldsymbol{R}_{kernel}$

This gives the disorientation matrix ($\boldsymbol{\omega_{ij}}$) for all of the kernel voxels in the central voxel reference frame. Next, the infinitesimal lattice rotation components is approximated by: $w_1 = (\omega_{32} - \omega_{23})/2$ , $w_2 = (\omega_{13} - \omega_{31})/2, w_3 = (\omega_{21} - \omega_{12})/2$ which are the components of the infinitesimal rotation vector $\boldsymbol{w}$. A result from a single slice of the measured volume is shown in Fig. 2-a.

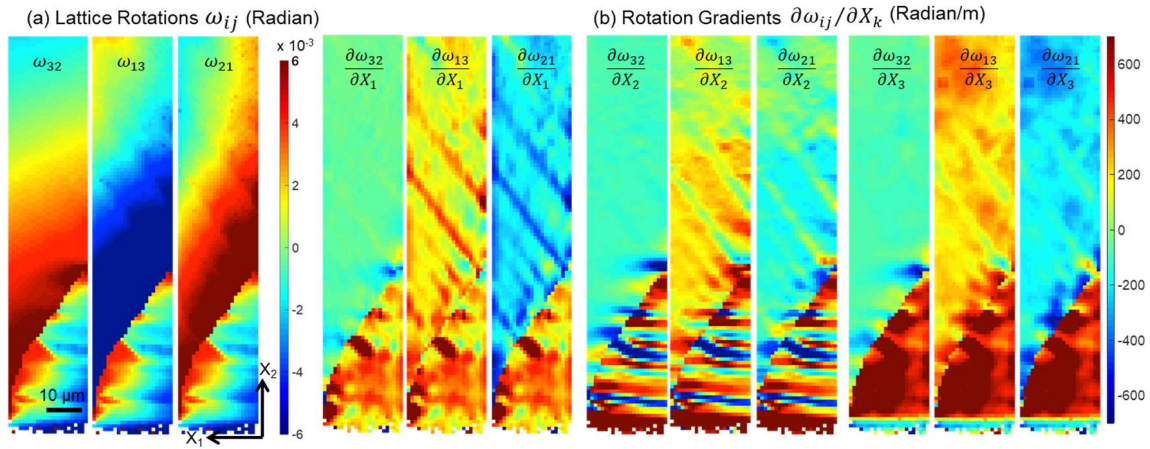

*Figure 2: Lattice rotation (a) and corresponding rotation gradient along 3 principal directions (b).*

The rotation gradient in all three directions in the kernel can be found by fitting the kernel to a hyper plane:

$$(5) \begin{pmatrix} w^1 \\ \vdots \\ w^n \end{pmatrix} = \begin{pmatrix} X_1^1 & X_2^1 & X_3^1 & 1 \\ \vdots & \vdots & \vdots & \vdots \\ X_1^n & X_2^n & X_3^n & 1 \end{pmatrix} \begin{pmatrix} g_{X_1} \\ g_{X_2} \\ g_{X_3} \\ C \end{pmatrix}$$

Where $g_{X_1} = \frac{\partial w}{\partial X_1}$ $g_{X_2} = \frac{\partial w}{\partial X_2}$ $g_z = \frac{\partial w}{\partial X_3}$ and $C$ is a constant.

In which $w^n$ refers to the infinitesimal rotation vector of the n[th] kernel voxel and $X_1^n$ $X_2^n$ and $X_3^n$ are its associated spatial coordinates. The calculated lattice curvature components are shown in Fig. 2-b.

A similar procedure is applied to obtain the strain gradients and the results are shown in Fig. 3-b together with the deviatoric strains (Fig. 3-a).

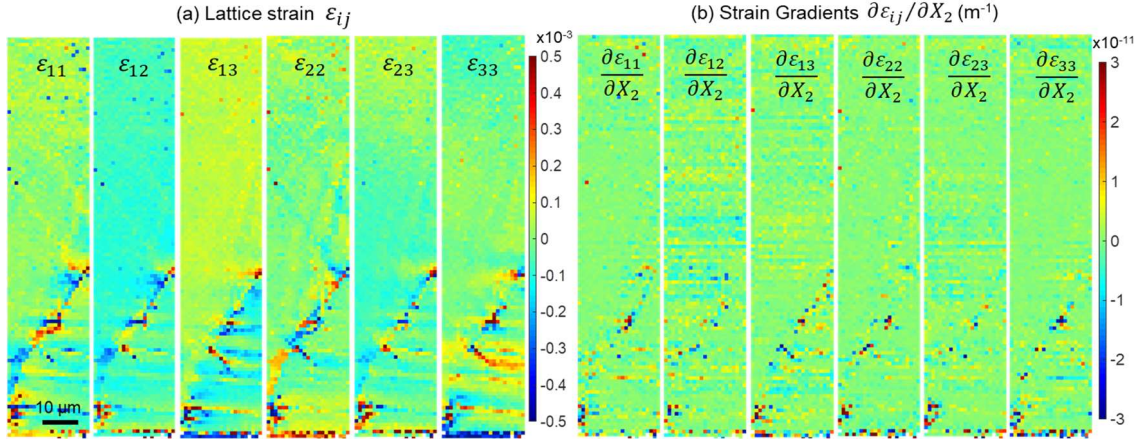

Figure 3: Elastic strain (a) and the corresponding strain gradient along Y direction.

As can be seen, for this case the magnitude of the elastic strain gradient is negligible compared to that of the rotation gradient, consistent with other studies on a more severe plastic deformation [40]. Therefore, we can ignore the elastic strain part ($\alpha_{ij}^\varepsilon$) of equation 1 and link the rotational part ($\alpha_{ij}^\omega$) to a dislocation tensor:

$$(6) \quad \begin{pmatrix} b_1^1 l_1^1 & b_1^2 l_1^2 & b_1^3 l_1^3 & & b_1^k l_1^k \\ b_1^1 l_2^1 & b_1^2 l_2^2 & b_1^3 l_2^3 & & b_1^k l_2^k \\ b_1^1 l_3^1 & b_1^2 l_3^2 & b_1^3 l_3^3 & & b_1^k l_3^k \\ b_2^1 l_1^1 & b_2^2 l_1^2 & b_2^3 l_1^3 & & b_2^k l_1^k \\ b_2^1 l_2^1 & b_2^2 l_2^2 & b_2^3 l_2^3 & \ldots & b_2^k l_2^k \\ b_2^1 l_3^1 & b_2^2 l_3^2 & b_2^3 l_3^3 & & b_2^k l_3^k \\ b_3^1 l_1^1 & b_3^2 l_1^2 & b_3^3 l_1^3 & & b_3^k l_1^k \\ b_3^1 l_2^1 & b_3^2 l_2^2 & b_3^3 l_2^3 & \ldots & b_3^k l_2^k \\ b_3^1 l_3^1 & b_3^2 l_3^2 & b_3^3 l_3^3 & & b_3^k l_3^k \end{pmatrix} \begin{pmatrix} \rho^1 \\ \rho^2 \\ \rho^3 \\ \vdots \\ \rho^k \end{pmatrix} = \begin{pmatrix} \frac{\partial \omega 12}{\partial x3} - \frac{\partial \omega 13}{\partial x2} \\ \frac{\partial \omega 13}{\partial x1} \\ \frac{\partial \omega 21}{\partial x1} \\ \frac{\partial \omega 32}{\partial x2} \\ \frac{\partial \omega 23}{\partial x1} - \frac{\partial \omega 21}{\partial x3} \\ \frac{\partial \omega 21}{\partial x2} \\ \frac{\partial \omega 32}{\partial x3} \\ \frac{\partial \omega 13}{\partial x3} \\ \frac{\partial \omega 31}{\partial x2} - \frac{\partial \omega 32}{\partial x1} \end{pmatrix}$$

Where $\rho^k$ is the density of the $k$th type of dislocation and $b_i^k$ and $l_j^k$ are the components of the Burgers vector and line direction of the $k$th type of dislocation. The total dislocation density is then $\sum_1^k \rho^k$. Nye analysed dislocations in a simple cubic system, where $b_i^k l_j^k$ have one to one correspondence with each of the nine lattice curvature terms, i.e. the diagonal part of the $\alpha_{ij}^\omega$ tensor represent a twist of the crystal axis due to the presence of the 3 screw dislocations and the off-axis parts correspond to a bending of the crystal faces due to the edge dislocations. In a typical crystal, there are usually more than nine dislocation types, for example there are 33 types of frequently referenced slip systems in *α*-titanium [41], rendering equation (4) under-determined. This problem can be treated by utilising optimisation methods, such as L[1] that minimises the total line energy [42] or L[2] that minimise the sum of squares of dislocation length. The L[1] optimisation is used here as this approach has physical meaning as dislocations can be considered in a low energy elastic energy configuration (considering their self-energy only) [43]. The L[1] optimisation is implemented in MATLAB using the 'linprog' algorithm as well as a weighting factor [42]:

(7) $\frac{E_{edge}}{E_{screw}} = \frac{1}{(1-\nu)}$

where $\nu$ is Poisson's ratio. Throughout the analysis the basis set of dislocation types are chosen to be pure screw and pure edge types. Where mixed dislocations are present in the crystal this will be captured in the results as a proportionate mix of edge and screw dislocation density for the given slip system. These fully 3D measurements can be downgraded to mimic information available from 2D HR-EBSD studies, for which the gradients in the $x_3$ direction are not accessible and equation (6) is truncated to remove the corresponding gradient terms together with the $b_3^k l_j^k$ terms of the dislocation tensor (see equation (2) in reference [41]). Furthermore, in working with 2D data, El-Dasher et al. [18] suggested using the entrywise 1-norm of the rotation gradient tensor ($\alpha_{ij}^\omega$) to provide a simple single scalar estimate of the total GND density:

(8) $\rho_{total} = \sum_1^i \sum_1^j |\alpha_{ij}^\omega|$

Fig. 4 compares dislocation densities estimated from the Nye-Kroner-Bilby (NKB) solutions (equation (6)), and the entrywise 1-norm (equation (8)) using different numbers of lattice curvature terms representing to 3D (9 terms) and 2D (<=6 terms) measurements.

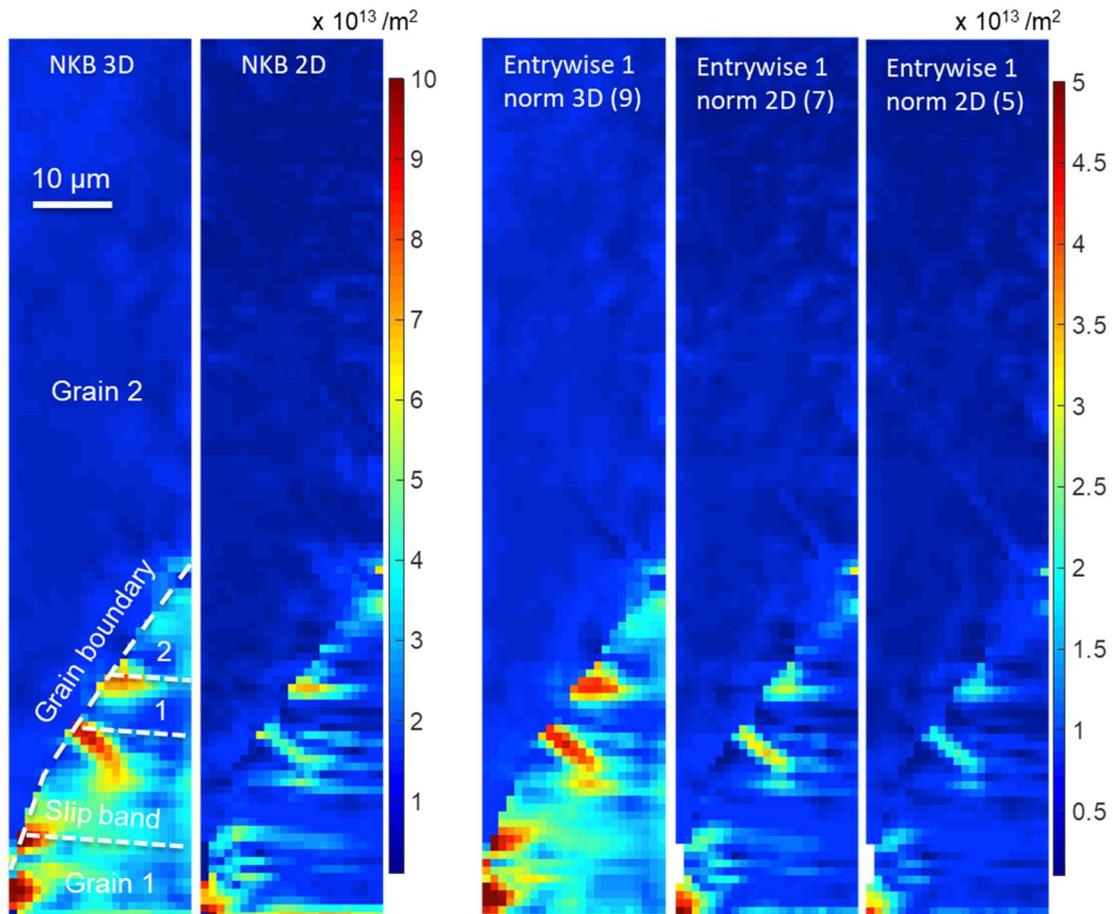

*Figure 4: Comparison of dislocation density distribution from solving equation (7) using all lattice curvature terms (Nye-Kroner-Bilby, NKB 3D), 6 lattice curvature terms (NKB 2D), by solving equation (9) using all lattice curvature terms (entrywise 1-norm 3D (9)), 7 terms (entrywise 1-norm 2D (7)), and 5 terms (entrywise 1-norm 2D (5)).*

It can be seen from Fig. 4 that discrete GND density 'hot spots' can be observed at the grain boundary plane, each of which corresponds to an intersection between a slip band and the grain boundary. High GND concentrations (>$10^{13}$ dislocations per $m^2$) in grain 1 can be found within 8 µm distance away from the grain boundary and around the slip plane-grain boundary intersections. The GNDs in grain 1, the slipping grain, are higher in extent compared to those in grain 2, the (elastically) sheared grain. It is directly evident from Fig. 4 that using fewer lattice curvature terms gives a lower GND magnitude but the spatial concentration of GNDs is sharper. This is quantitatively reflected by the 1D line-trace analysis in Fig.5, from which it can be seen that the entrywise 1-norm is an underestimation of the GND densities

Page | 12

(Fig. 5-a) with the magnitude systematically dropping with fewer lattice curvature terms entering the calculation (i.e. the system of equations is less constrained). On the other hand, the peak width is similar between the NKB-3D and the 9-terms entrywise 1-norm but is smaller for the 2D calculations, indicating a sharper spatial concentration of the GND density. It seems that the 2D calculations (i.e. with <= 6 lattice curvature terms) are sufficiently representative of the dislocation density distribution with the added benefit of sharper peaks to aid identification of significant deformation concentrators (in this case, slip band-grain boundary interactions).

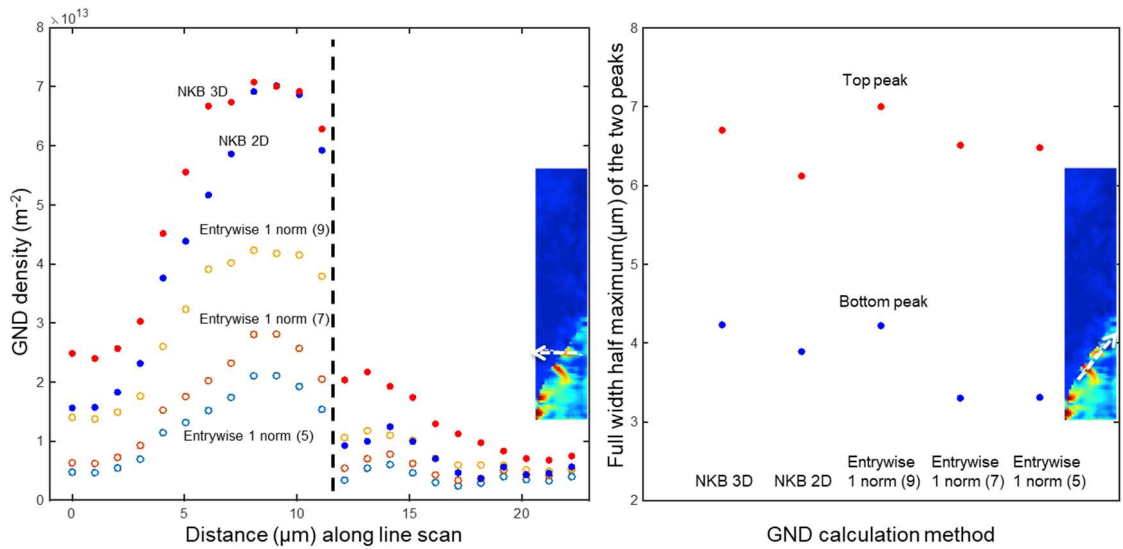

*Figure 5: (a) line scan along one of the GND hot spot in the direction indicated by the white arrow. (b) GND peak breadth of the peak profile from the indicated line scan along grain boundary across two GND hot spots.*

The entrywise 1-norm of the lattice curvature tensor (equation (8)) provides a solution to the NKB equation (equation (6)), providing a quick solution to the GND density. However, it is impossible to consider the densities of individual types of dislocations giving rise to the measured lattice curvatures. The line energy minimisation approach was found to provide a good approximation of individual dislocation densities by comparison with ECCI [44]. Applying a similar approach in the current investigation determines that the GND density hotspot in the slipping grain (grain 1) is mainly <a> pyramidal edge dislocations together with two variants of screw dislocations. The results are shown in Fig. 6.

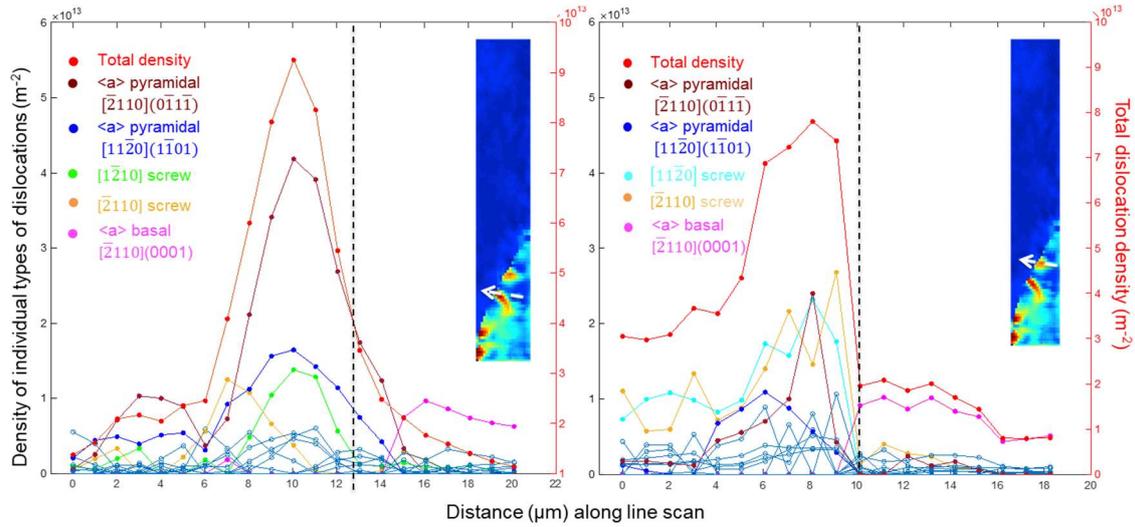

*Figure 6: Line profiles along 2 selected slip bands, in the slip band direction across the grain boundary. Indicating individual GND densities from solving equation (7).*

The apparent drop of GND density near the grain boundary, in Fig. 5 and Fig. 6, is likely due to the drop of indexing quality, resulting from diffraction pattern overlap at the grain boundary.

## 4 Discussion

We have presented the GND density distribution due to the local plasticity as a result of arrays of dislocation lines piled up against a grain boundary. The 3D resolution of the curvature field with the DAXM method provides sufficient spatial resolution (< 0.5 μm [28,45], smaller than the scanning step size) and high angular resolution (~0.01°) for 3D mapping of the curvature tensor. A lower bound GND density noise level can be estimated using:

(9) $\Delta \rho = \frac{\delta}{b\lambda}$

Where $\delta$ is the angular resolution (in radians) of the measurement technique, $b$ is the Burgers vector magnitude, and $\lambda$ is the step size [42]. Using $b$=0.295 nm the GND noise level is estimated to be ≈6x10$^{11}$ m$^{-2}$ for DAXM at a step size of 1 μm, and therefore the GND densities presented are sufficiently above the noise floor for the results to be valid. We note that the measurement step size is important, and here the step size needs to be sufficient to measure the net curvatures around the features of interest and related to the stored GND content (for a more complete investigation of the effect of step size on experimentally measured GND density please see Jiang et al [26]).

The distribution of dislocation densities is heterogeneous with a significantly higher magnitude and extent of the distribution in the slipping grain. They are also mainly confined to regions close to the slip band – grain boundary intersections. If sufficiently high, the elastic stress field of the dislocation pile up will operate a regenerative dislocation source [46]. Here, this is observed close to the grain boundary and results in dislocations confined to the neighbourhood of the pile-up front, as is evident in the observed GND density concentrations in Fig. 4, with the exception of the one which appears to show dislocations that are 'reflected' back into grain 1 along certain crystallographic directions (indicated by dashed circle in Fig. 7). This GND field is not due to the pile up of other slip planes as the $[2\bar{1}\bar{1}0]$ $(01\bar{1}0)$ prismatic slip is the only deformation system in grain 1. This is evidenced by the elastic strain map shown in Fig. 3 together with a previous surface EBSD analysis. A slip trace analysis, Fig.7, demonstrates that this GND field is localised along a $(01\bar{1}1)$ pyramidal slip plane. Furthermore, Schmid factor analysis, Table 1, indicates that there are in fact two variants of the <a> type $\{01\bar{1}1\}$ pyramidal slip system with high Schmid factors (using the macroscopic loading direction during the deformation of the polycrystalline sample). However, while the macroscopic Schmid factor indicates how the global stress state is resolved, it is less likely to be the only reason for the observed feature due to the absence of similar features for other GND density fields. There is also an absence of a long range GND density field along the prismatic slip plane traces, whose activation require lower stresses compared to the pyramidal slip systems [47]. Li [30] proposed grain boundary ledge as a dislocation source with its theory developed by Hirth [48], calling these sources disconnections[†]. Both analyses imply a directionality of the dislocations emitted from this type of grain boundary source, with Li's theory Schmid factor independent and Hirth's theory Schmid factor dependent.

---

[†] We will use the term 'ledge' in this paper to emphasize the geometric nature of this investigation.

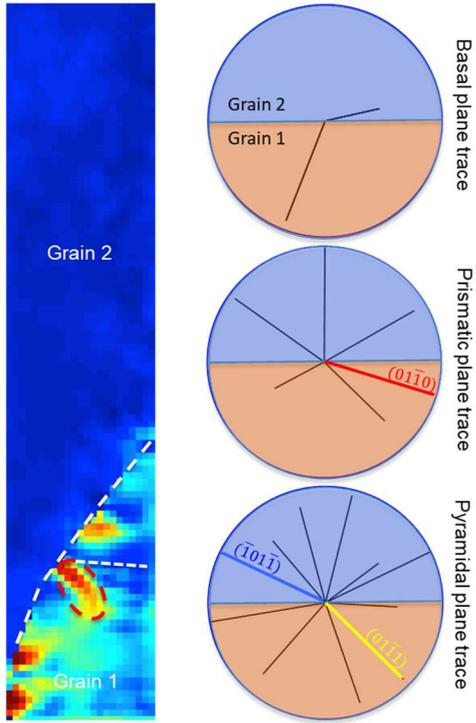

*Figure 7: Slip traces of basal, prismatic, and pyramidal planes, with traces of grain 1 plotted in the southern hemisphere and those of grain 2 in the northern hemisphere.*

| (a) | Schmid factor of slip systems in grain 1 | | | | | | | | | | | |
|---|---|---|---|---|---|---|---|---|---|---|---|---|
| <a> Basal | [11$\bar{2}$0] (0001) | [1$\bar{2}$10] (0001) | [$\bar{2}$110] (0001) | | | | | | | | | |
| | 0.04 | 0.01 | 0.02 | | | | | | | | | |
| <a> Prismatic | [2$\bar{1}$$\bar{1}$0] (01$\bar{1}$0) | [11$\bar{2}$0] ($\bar{1}$100) | [1$\bar{2}$10] ($\bar{1}$010) | | | | | | | | | |
| | 0.48 | 0.12 | 0.36 | | | | | | | | | |
| <a> Pyramidal | [2$\bar{1}$$\bar{1}$0] (01$\bar{1}$1) | [11$\bar{2}$0] ($\bar{1}$101) | [1$\bar{2}$10] ($\bar{1}$011) | [1$\bar{2}$10] ($\bar{1}$011) | [11$\bar{2}$0] ($\bar{1}$10$\bar{1}$) | [2$\bar{1}$$\bar{1}$0] (0$\bar{1}$11) | | | | | | |
| | 0.43 | 0.09 | 0.32 | 0.31 | 0.12 | 0.41 | | | | | | |
| <c+a> Pyramidal | [2$\bar{1}$$\bar{1}$3] ($\bar{1}$101) | [1$\bar{2}$1$\bar{3}$] ($\bar{1}$101) | [2$\bar{1}$$\bar{1}$3] ($\bar{1}$01$\bar{1}$) | [$\bar{1}$$\bar{1}$23] ($\bar{1}$01$\bar{1}$) | [2$\bar{1}$$\bar{1}$3] ($\bar{1}$011) | [11$\bar{2}$3] ($\bar{1}$011) | [2$\bar{1}$$\bar{1}$$\bar{3}$] ($\bar{1}$10$\bar{1}$) | [1$\bar{2}$1$\bar{3}$] ($\bar{1}$10$\bar{1}$) | [11$\bar{2}$3] (0$\bar{1}$11) | [1$\bar{2}$1$\bar{3}$] (0$\bar{1}$11) | [$\bar{1}$$\bar{1}$23] (01$\bar{1}$1) | [$\bar{1}$2$\bar{1}$$\bar{3}$] (01$\bar{1}$1) |
| | 0.03 | 0.02 | 0.24 | 0.41 | 0.28 | 0.44 | 0.04 | 0.03 | 0.38 | 0.16 | 0.36 | 0.13 |
| (b) | Schmid factor of slip systems in grain 2 | | | | | | | | | | | |
| <a> Basal | [11$\bar{2}$0] (0001) | [1$\bar{2}$10] (0001) | [$\bar{2}$110] (0001) | | | | | | | | | |
| | 0.29 | 0.08 | 0.21 | | | | | | | | | |
| <a> Prismatic | [2$\bar{1}$$\bar{1}$0] (01$\bar{1}$0) | [11$\bar{2}$0] ($\bar{1}$100) | [1$\bar{2}$10] ($\bar{1}$010) | | | | | | | | | |
| | 0.05 | 0.02 | 0.02 | | | | | | | | | |
| <a> Pyramidal | [2$\bar{1}$$\bar{1}$0] (01$\bar{1}$1) | [11$\bar{2}$0] ($\bar{1}$101) | [1$\bar{2}$10] ($\bar{1}$01$\bar{1}$) | [1$\bar{2}$10] ($\bar{1}$011) | [11$\bar{2}$0] ($\bar{1}$10$\bar{1}$) | [2$\bar{1}$$\bar{1}$0] (0$\bar{1}$11) | | | | | | |
| | 0.14 | 0.12 | 0.06 | 0.02 | 0.16 | 0.06 | | | | | | |
| <c+a> Pyramidal | [2$\bar{1}$$\bar{1}$3] ($\bar{1}$101) | [1$\bar{2}$1$\bar{3}$] ($\bar{1}$101) | [2$\bar{1}$$\bar{1}$3] ($\bar{1}$01$\bar{1}$) | [$\bar{1}$$\bar{1}$23] ($\bar{1}$01$\bar{1}$) | [2$\bar{1}$$\bar{1}$3] ($\bar{1}$011) | [11$\bar{2}$3] ($\bar{1}$011) | [2$\bar{1}$$\bar{1}$$\bar{3}$] ($\bar{1}$10$\bar{1}$) | [1$\bar{2}$1$\bar{3}$] ($\bar{1}$10$\bar{1}$) | [11$\bar{2}$3] (0$\bar{1}$11) | [1$\bar{2}$1$\bar{3}$] (0$\bar{1}$11) | [$\bar{1}$$\bar{1}$23] (01$\bar{1}$1) | [$\bar{1}$2$\bar{1}$$\bar{3}$] (01$\bar{1}$1) |
| | 0.35 | 0.29 | 0.49 | 0.46 | 0.18 | 0.18 | 0.36 | 0.45 | 0.25 | 0.22 | 0.42 | 0.49 |

*Table 1: Schmid factor analysis of various slip systems in grain 1 (a) and grain 2 (b).*

To investigate the crystallographic nature of the grain boundary ledge due to the slip band-grain boundary interaction, it is important to understand the shear transfer across the grain boundary. Trace analysis in Fig. 7 indicates the active dislocation pile up in grain 1 is on the $[2\bar{1}\bar{1}0]$ $(01\bar{1}0)$ prismatic slip system. We define two sets of cubic reference frames. For the DAXM experiment, the crystal reference frame is defined where $X_{crystal}$ is parallel to the $[2\bar{1}\bar{1}0]$ Burgers vector direction (i.e. [100] direction in Miller indices), $Z_{crystal}$ parallel to [0001] (i.e. [001]) direction, and $Y_{crystal}$ perpendicular to the plane defined by the $X_{crystal}$ and $Y_{crystal}$ directions, i.e. [120] or $(01\bar{1}0)$ plane normal direction. The slip reference frame is defined such that $X_{slip}$ is parallel to $[2\bar{1}\bar{1}0]$, $Z_{slip}$ parallel to $[01\bar{1}0]$, and $Y_{slip}$ parallel to [0001]. The slip shear, in slip reference frame, takes the following form:

(10) $\varepsilon^{slip} = \begin{pmatrix} 0 & 0 & 1 \\ 0 & 0 & 0 \\ 0 & 0 & 0 \end{pmatrix}$

The $\varepsilon^{slip}$ is resolved onto the cubic crystal reference frame using a suitable structure matrix taking the following form [49]:

(11)
$$HC = \begin{pmatrix} a & b*\cos\gamma & c*\cos\beta \\ 0 & b*\sin\gamma & c*(\cos\alpha - \cos\beta\cos\gamma)/\sin\gamma \\ 0 & 0 & c*(1+2\cos\alpha\cos\beta\cos\gamma - \cos^2\alpha - \cos^2\beta - \cos^2\gamma)^2/\sin\gamma \end{pmatrix}$$

Where a, b, c, α, β, γ, are lattice constants and HC indicate switching reference frame from 'Hexagonal to cubic'.

And a transformation matrix can be defined:

(12) $HCR = [HC * \begin{bmatrix} b(u) \\ b(v) \\ b(w) \end{bmatrix}, HC * \begin{bmatrix} l(u) \\ l(v) \\ l(w) \end{bmatrix}, (HC^{-1})^T * \begin{bmatrix} n(h) \\ n(k) \\ n(l) \end{bmatrix}]$

Where *b* is Burgers vector direction, *n* is slip plane. The (*u v w*), and (*h k l*) are the Miller indices of the burgers vector and slip plane respectively. The *l* is the direction perpendicular to the plane defined by *b* and *n* and is equal to the cross product between the first and the third column of the HCR matrix.

Slip shear in crystal reference frame is obtained using:

(13) $\varepsilon^{crystal} = HCR^{-1} * \varepsilon^{slip} * HCR = \begin{pmatrix} 0 & -1 & 0 \\ 0 & 0 & 0 \\ 0 & 0 & 0 \end{pmatrix}$

The displacement gradient tensor, associated with this slip, can be assessed in the crystal reference frame of grain2 using:

(14) $\Delta g = g^2(\varphi_1^2\ \phi^2\ \varphi_1^2) \cdot g^1(\varphi_1^1\ \phi^1\ \varphi_1^1)^T$

(15) $\beta^{grain2} = \Delta g \cdot \varepsilon^{crystal} \cdot \Delta g^T = \begin{pmatrix} 0.271 & -0.122 & -0.066 \\ 0.146 & -0.066 & -0.036 \\ 0.835 & -0.377 & -0.205 \end{pmatrix}$

where $g$ is the rotation matrix and $(\varphi_1\ \phi\ \varphi_2)$ are the Euler angles (grain1: (268.56 73.29 283.65), grain2: (114.27 1.27 313.73)). A step-by-step guide of the above calculation can be found in appendix A1.

For $\beta^{grain2}$, $|\beta_{12}^{grain}| + |\beta_{21}^{grain2}| = 0.268$ indicate the level of shear on <a> prismatic plane in grain 2 as a result of the shear in grain 1, $|\beta_{13}^{grai}| + |\beta_{23}^{grain2}| = 0.102$ on <a> basal plane, and $|\beta_{31}^{grai}| + |\beta_{32}^{grai}| = 1.212$ on the <c+a> pyramidal plane. These are shown graphically in Fig. 8.

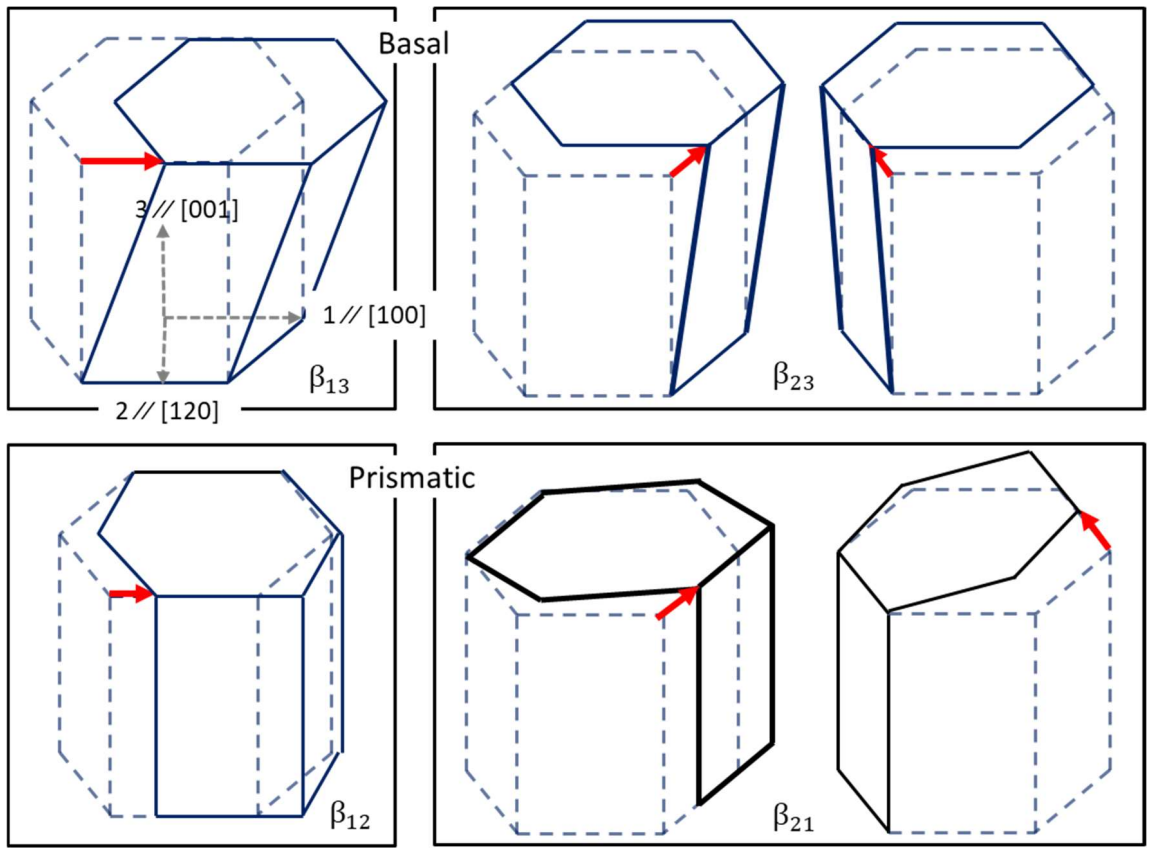

*Figure 8: Schematic representation of the relationship between displacement gradient tensor components and the related physical shape change (The origin of the analysis is from [50]). Crystal reference frame is defined in the first image.*

It can be seen from the above analysis that the shearing of grain 1 is likely to be accommodated by a <c+a> pyramidal type deformation system in grain 2. To find this deformation system in grain 2, a geometrical factor [51] was assessed between the active prismatic slip system in grain 1 to slip systems in grain 2 and results shown in Table 2-a. The geometrical factor takes the form of $M = |(\hat{b}_i \cdot \hat{b}_j)(\hat{n}_i \cdot \hat{n}_j)|$, where $\hat{b}$ and $\hat{n}$ are the unit burgers vector and slip plane normal vectors respectively. The *M* ranges from 0 to 1 with values approaching 1 represent perfect alignment where the shear transfer is the most effective. The shear transfer is physically carried by the transfer of Burgers vectors across the grain boundary, during which the grain boundary is displaced proportionally to the number of transfer events. Hence, a grain boundary ledge is created together with the residual Burgers vector to accommodate the Burgers vector mismatch. This is schematically shown in Fig. 9-a.

| (a) | M1-2 (prismatic slip in grain 1 to slip systems in grain 2) | | | | | | | | | | | |
|---|---|---|---|---|---|---|---|---|---|---|---|---|
| <a> Basal | [11$\bar{2}$0] (0001) | [1$\bar{2}$10] (0001) | [$\bar{2}$110] (0001) | | | | | | | | | |
| | 0.28 | 0.02 | 0.3 | | | | | | | | | |
| <a> Prismatic | [2$\bar{1}\bar{1}$0] (01$\bar{1}$0) | [11$\bar{2}$0] ($\bar{1}$100) | [1$\bar{2}$10] ($\bar{1}$010) | | | | | | | | | |
| | 0.27 | 0.15 | 0.02 | | | | | | | | | |
| <a> Pyramidal | [2$\bar{1}\bar{1}$0] (01$\bar{1}$1) | [11$\bar{2}$0] ($\bar{1}$101) | [1$\bar{2}$10] ($\bar{1}$01$\bar{1}$) | [1$\bar{2}$10] ($\bar{1}$011) | [11$\bar{2}$0] ($\bar{1}$10$\bar{1}$) | [2$\bar{1}\bar{1}$0] (0$\bar{1}$11) | | | | | | |
| | 0.37 | 0.01 | 0.03 | 0.01 | 0.26 | 0.09 | | | | | | |
| <c+a> Pyramidal | [2$\bar{1}\bar{1}$3] ($\bar{1}$101) | [$\bar{1}$2$\bar{1}$3] ($\bar{1}$101) | [2$\bar{1}\bar{1}\bar{3}$] ($\bar{1}$01$\bar{1}$) | [$\bar{1}$1 2 3] ($\bar{1}$01$\bar{1}$) | [2$\bar{1}\bar{1}\bar{3}$] ($\bar{1}$011) | [$\bar{1}$1$\bar{2}$3] ($\bar{1}$011) | [2$\bar{1}\bar{1}$3] ($\bar{1}$10$\bar{1}$) | [$\bar{1}$2$\bar{1}$3] ($\bar{1}$10$\bar{1}$) | [$\bar{1}$1$\bar{2}$3] (0$\bar{1}$11) | [$\bar{1}$2$\bar{1}\bar{3}$] (0$\bar{1}$11) | [$\bar{1}$1 2 3] (0$\bar{1}$11) | [$\bar{1}$2$\bar{1}$3] (0$\bar{1}$11) |
| | 0.01 | 0.01 | **0.95** | 0.94 | 0.18 | 0.19 | 0.52 | 0.38 | 0.08 | 0.12 | 0.69 | 0.49 |
| Contraction twin | [10$\bar{1}\bar{2}$] (10-11) | [$\bar{1}$102] (1-10-1) | [01$\bar{1}\bar{2}$] (01-11) | [$\bar{1}$012] (-1011) | [$\bar{1}$102] (-110-1) | [0$\bar{1}$12] (0-111) | | | | | | |
| | **0.98** | 0.01 | 0.61 | 0.19 | 0.46 | 0.10 | | | | | | |
| (b) | M2-1 (<c+a> pyramidal slip in grain 2 to slip systems in grain 1) | | | | | | | | | | | |
| <a> Basal | [11$\bar{2}$0] (0001) | [1$\bar{2}$10] (0001) | [$\bar{2}$110] (0001) | | | | | | | | | |
| | 0.07 | 0.04 | 0.12 | | | | | | | | | |
| <a> Prismatic | [2$\bar{1}\bar{1}$0] (01$\bar{1}$0) | [11$\bar{2}$0] ($\bar{1}$100) | [1$\bar{2}$10] ($\bar{1}$010) | | | | | | | | | |
| | **0.95** | 0.36 | 0.14 | | | | | | | | | |
| <a> Pyramidal | [2$\bar{1}\bar{1}$0] (01$\bar{1}$1) | [11$\bar{2}$0] ($\bar{1}$101) | [1$\bar{2}$10] ($\bar{1}$01$\bar{1}$) | [1$\bar{2}$10] ($\bar{1}$011) | [11$\bar{2}$0] ($\bar{1}$10$\bar{1}$) | [2$\bar{1}\bar{1}$0] (0$\bar{1}$11) | | | | | | |
| | **0.89** | 0.35 | 0.14 | 0.10 | 0.28 | 0.78 | | | | | | |
| <c+a> Pyramidal | [2$\bar{1}\bar{1}$3] ($\bar{1}$101) | [$\bar{1}$2$\bar{1}$3] ($\bar{1}$101) | [2$\bar{1}\bar{1}\bar{3}$] ($\bar{1}$01$\bar{1}$) | [$\bar{1}$1 2 3] ($\bar{1}$01$\bar{1}$) | [2$\bar{1}\bar{1}\bar{3}$] ($\bar{1}$011) | [$\bar{1}$1$\bar{2}$3] ($\bar{1}$011) | [2$\bar{1}\bar{1}$3] ($\bar{1}$10$\bar{1}$) | [$\bar{1}$2$\bar{1}$3] ($\bar{1}$10$\bar{1}$) | [$\bar{1}$1$\bar{2}$3] (0$\bar{1}$11) | [$\bar{1}$2$\bar{1}\bar{3}$] (0$\bar{1}$11) | [$\bar{1}$1 2 3] (0$\bar{1}$11) | [$\bar{1}$2$\bar{1}$3] (0$\bar{1}$11) |
| | 0.19 | 0 | 0.28 | 0.20 | 0.09 | 0.04 | 0.33 | 0.18 | 0.11 | 0.31 | 0.47 | 0 |

Table 2: (a) Geometric alignment between the activated prismatic slip system in grain 1 and various slip and a twin system in grain 2. (b) the alignment between maximum sheared twin variant in grain 2 to various slip systems in grain 1.

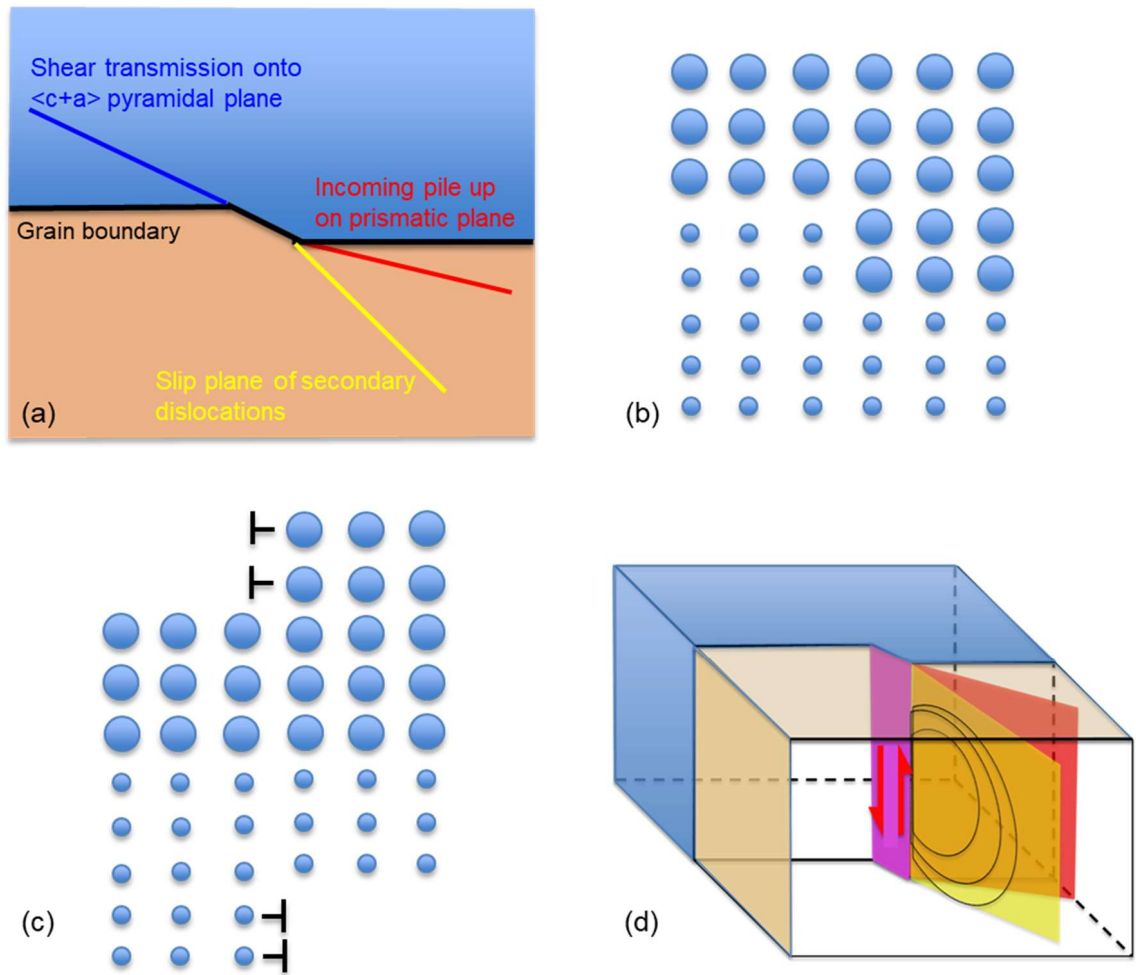

Figure 9: (a) Schematic arrangement of the grain boundary ledge formed by a dislocation pile up (red), the direct shear transfer plane (blue), and the secondary dislocation emission plane (yellow). (b) The local atomic arrangement at a ledge. (c) Edge dislocation emission by Li mechanism [30]. (d) Screw dislocation emission by Hirth mechanism [31].

The ledge is formed where the prismatic plane in grain 1 is forced to match the pyramidal plane in grain 2. A contraction twin of the variant $[10\bar{1}2]$ $(10\bar{1}1)$ could nucleate with the condition that the shear on the active slip system is sustained and that the c-axis of the grain 2 crystal is compressed, and this is similar to slip assisted twin nucleation which has been previously observed [52], while the formation of the [2$\bar{1}\bar{1}$3] (10$\bar{1}$1) slip system may be possible at higher strain due to the higher critical resolved shear stress [53]. As the sample was unloaded at 1% strain, the critical condition is met neither for the twin nor the <c+a> pyramidal slip, formerly rendering the ledge energetically metastable. Local atomic relaxation could lead to ledge annihilation and emission of edge dislocations in the adjacent grains (Fig.9-b, c). The emitted dislocations would directly travel back along the prismatic slip trace in grain 1 as they are near perfectly aligned. However, this is not energetically favourable due to a repulsive stress from the existing dislocations in the slip plane. Among the other slip systems in grain 1, the [2$\bar{1}\bar{1}$0] (01$\bar{1}$1) pyramidal slip variant has a very close $M$ level compared to the prismatic slip variant. It is possible that the emitted dislocations are transferred onto the pyramidal slip plane, this process will also require the generation of residual dislocations of opposite sign as those created during the ledge formation and hence cancel each other out. This pyramidal slip would produce a slip trace in the direction of the observed GND density distribution and is consistent with the slip trace analysis in Fig. 7. Note that the [2$\bar{1}\bar{1}$0] (01$\bar{1}$1) pyramidal slip trace identified from the geometrical analysis is consistent with the variant dominating the local lattice curvature identified through solving the NKB equation (Fig. 6).

This mechanism has been observed before, where secondary dislocations reflected back into the slipping grain at grain boundary ledges are evident in SEM studies such as Fig. 4 in [51] and Fig. 8 in [25]. Sub-micron morphologies of grain boundary ledges were also shown by Murr [54] using TEM observations. Other TEM investigations [55,56] have shown secondary dislocation emission from grain boundary features closely resembling the ledges described by Murr, and in some cases the gain boundary ledge could lead to multiple slip variants in the slipping grain, see Fig. 5 in reference [57]. The emission effect was attributed to the minimisation of grain boundary energy density and is implicitly consistent with the source annihilation mechanism by Li [30]. However, the capacity of dislocation emission due to source annihilation may be limited. This is due to the number of emitted dislocations dependent on the ledge height which is determined by the number of dislocations incorporated into the grain boundary [58]. A more potent mechanism would be for the ledge to emit screw dislocations, due to local shearing along the ledge plane, as described by Price and Hirth [59] and shown schematically in Fig. 9-d. The emitted screw dislocations have their dragging force increased due to the accompanying emission of grain boundary dislocations, however, they may be glissile; moving via double cross slip and/or they may multiply through the Frank-Read mechanism. This is more likely a mechanism that gives rise to the far-reaching GND density field which is absent in the neighbouring grain (for the former has more favourable slip condition, i.e. Schmid factor and <a> pyramidal slip (Table 1) in grain 1, while dislocation glide on <c+a> slip system in grain 2 is energetically more expensive).

It is noted that under certain circumstances, such as at twin boundaries, screw dislocations can transfer across a grain boundary by cross slip without creating a ledge [60]. Furthermore, ledges may also form in the slipping grain without the counter part in the adjacent grain, in which case, either GNDs or a void is necessary to accommodate the local shape change. In the present work, 5 slip planes of the same variant in a single grain intercept a common grain boundary at different locations, leading to local plasticity with differing density and distribution (this is consistent with the stress intensity ahead of each slip plane which varies against locations on the grain boundary plane, as observed in our prior work [35]). Factors such as oxygen distribution, internal dislocation structure of the grain boundary, and grain boundary topography all potentially contribute to the observed differences and could be potentially be homogenised into an 'energy fluctuation' term on top of the 5-parameter grain boundary character distribution, provided their individual interactions with dislocation pile-ups can be identified.

## 5. Conclusion

The advances of high spatial and high angular resolution X-ray micro-diffraction have made it possible to resolve spatially varying orientations at the micrometer scale. This has enabled us to visualise the GND density distribution due to micro-plasticity where arrays of dislocations were blocked by a grain boundary. Some notable observations are summarised below:

1. The GND density distribution obtained from a 2D measurement corroborates with a 3D GND density measurement. These distributions can also be revealed by entrywise 1-norm of the lattice curvature tensor without solving the Nye-Kröner-Bilby equation.
2. Dislocations mainly concentrate in the slipping grain if the neighbouring grain does not provide an easy shear path.
3. Slip band-grain boundary interactions can lead to grain boundary ledge formation, which can be used to explain the phenomena of secondary dislocation emission from the grain boundary.
4. The crystallographic character of the grain boundary ledge as well as the direction of the secondary dislocation emission can be determined by assessing the geometrical alignment of the shear path across a grain boundary.
5. Both the magnitude and distribution of GND densities near a grain boundary are different at various locations of the grain boundary plane, even though the dislocation pileups that created such GND density concentrations are of the same variant.

**Author Contributions**

YG conducted analysis of the experimental data, developed the model proposed in Figure 9 and drafted the final manuscript. FH assisted with the DAXM data analysis. TBB, ET and DC conducted the measurements. TBB and AJW supervised the work. All authors revised the manuscript and agreed to the final draft.

**Acknowledgement**

Experimental work and the majority part of the analysis was conducted in University of Oxford funded by the HexMat programme grant (EP/K034332/1). The rest of the work was completed at Imperial College London under the support of AVIC-BIAM. TBB acknowledges funding of his research fellowship from the Royal Academy of Engineering. ET acknowledges support through an EPSRC Early Career Fellowship (EP/N007239/1). FH acknowledges funding from the European Research Council (ERC) under the European Union's Horizon 2020 research and innovation programme (grant agreement No 714697). We acknowledge the help from Dr. Wenjun Liu and Dr. Ruqing Xu for their help with the experiment, and we greatly appreciate the communication with Dr. Jon Tischler for interpretation and analysis of experimental results. This research used resources (34-ID-E) of the Advanced Photon Source, a U.S. Department of Energy (DOE) Office of Science User Facility operated for the DOE Office of Science by Argonne National Laboratory under Contract No. DE-AC02-06CH11357. We acknowledge Mr. Simon Wyatt for proof reading the manuscript.

Appendix A:

A1: Step-by-step guild from equation 11 to 13

Equation 11 was given by Young and Lytton [49] to express Miller indices, which are commonly referenced in crystal reference frame, to a cubic (Cartesian) reference frame. Equation 11 apply to general triclinic crystal system and in the case of titanium which has hexagonal close packed (HCP) crystal structure the lattice constants are: a=b=0.295 nm, c= 0.468 nm, α=β=90°, and γ=120°, the HC matrix becomes:

$$(A1.1)\ HC = \begin{pmatrix} 0.295 & -0.148 & 0 \\ 0 & 0.256 & 0 \\ 0 & 0 & 0.468 \end{pmatrix}$$

To transform directional vectors to cubic reference frame, the following is applied:

$$(A1.2)\ \begin{pmatrix} u \\ v \\ w \end{pmatrix}_{cubic} = HC * \begin{pmatrix} u \\ v \\ w \end{pmatrix}_{hcp}$$

However to find the normal vectors of crystal planes, extra measure needs to be taken. This is because for non-cubic crystal the plane indices (h k l) is generally not coincident with the [h k l] direction, so plane normal direction needs to be found in reciprocal space [49]. Young and Lytton [49] had given a rather complex procedure which is equal to taking the transpose of the inverse of the HC matrix and apply the following:

$$(A1.3)\ \begin{pmatrix} h \\ k \\ l \end{pmatrix}_{cubic} = (HC^{-1})^T * \begin{pmatrix} h \\ k \\ l \end{pmatrix}_{hcp}$$

This leads to the general form of the HCR matrix shown in equation 12.

In the specific application demonstrated in the manuscript, $X_{slip}$ // [1 0 0] (in Miller indices notation), $Z_{slip}$ // normal direction of (0 1 0), applying equations A1.2 and A1.3, we find the slip direction and slip plane norm directions in cubic reference frame as:

$$(A1.4)\ X_{cubic} = HC * \begin{pmatrix} 1 \\ 0 \\ 0 \end{pmatrix} = \begin{pmatrix} 0.295 \\ 0 \\ 0 \end{pmatrix} = \begin{pmatrix} 1 \\ 0 \\ 0 \end{pmatrix}$$

(A1.5) $Z_{cubic} = (HC^{-1})^T * \begin{pmatrix} 0 \\ 1 \\ 0 \end{pmatrix} = \begin{pmatrix} 0 \\ 3.914 \\ 0 \end{pmatrix} = \begin{pmatrix} 0 \\ 1 \\ 0 \end{pmatrix}$

(A1.6) $Y_{cubic} = Z_{cubic} \times X_{cubic} = \begin{pmatrix} 0 \\ 0 \\ -2.887 \end{pmatrix} = \begin{pmatrix} 0 \\ 0 \\ -1 \end{pmatrix}$

The above three equations give rise to a specific form of equation 12 as:

(A1.7) $HCR = \begin{pmatrix} 1 & 0 & 0 \\ 0 & 0 & 1 \\ 0 & -1 & 0 \end{pmatrix}$

The HCR matrix can be applied through equation 13 to assess tensors expressed in $X_{slip}$- $Y_{slip}$ reference frame in the $X_{crystal}$- $X_{crystal}$ reference. In this specific application:

(A1.8) $\varepsilon^{crystal} = HCR^{-1} * \begin{pmatrix} 0 & 0 & 1 \\ 0 & 0 & 0 \\ 0 & 0 & 0 \end{pmatrix} * HCR = \begin{pmatrix} 0 & -1 & 0 \\ 0 & 0 & 0 \\ 0 & 0 & 0 \end{pmatrix}$

A2: A video showing a 360° overview of the scanned volume. The quantity used to render the colour field is GND density. A voxel intensity thresholding was applied to filter out low intensity voxels and hence revealing the high intensity GND density field close to the grain boundary.

**References:**

[1] G.I. Taylor, Plastic strain in metals, J. Inst. Met. 62 (1938) 307–324.

[2] R. Becker, L.A. Lalli, Analysis of texture evolution in channel die compression Part II: Effects of Grains which Shear, Textures Microstruct. 14–18 (1991) 145–150. papers2://publication/uuid/CB6C7851-EAB6-42AE-B3BA-BA6D21DB58E1.

[3] S. V. Harren, H.E. Deve, R.J. Asaro, Overview Shear Band Formation in Plane, Acta Met. 36 (1988) 2435–2480.

[4] M.F. Ashby, The deformation of plastically non-homogeneous materials, Philos. Mag. 21 (1970) 399–424. doi:10.1080/14786437008238426.

[5] A. Arsenlis, D.M. Parks, Crystallographic aspects of geometrically necessary and statistically stored dislocation density, Acta Mater. 47 (1999) 1597–1611. doi:10.1016/S1359-6454(99)00020-8.

[6] J.F. Nye, Some Geometrical relations in dislocated crystals, Acta Mater. 1 (1953) 153–162.

[7] E. Kroner, Continuum theory of dislocations and self-stresses, Ergebnisse Der Angew. Math. 5 (1958) 1327–1347.

[8] B.A. Bilby, L.R.T. Gardner, E. Smith, The relationship between dislocation density and stress, Acta Metall. 6 (1958) 29–33.

[9] E.O. Hall, The deformation and aging of mild steel, Proc. Phys. Soc. B. 64 (1951) 747.

[10] N.J. Petch, The cleavage strength of polycrystals, J Iron Steel Inst. 25 (1953) 174.

[11] S. Sun, B.L. Adams, W.E. King, Observations of lattice curvature near the interface of a deformed aluminium bicrystal, Phil. Mag. A. 80 (2000) 9–25.

[12] W.W. Gerberich, N.I. Tymiak, J.C. Grunlan, M.F. Horstemeyer, M.I. Baskes, Interpretations of indentation size effects, J. Appl. Mech. . 69 (2002) 433–442.

[13] E. Demir, D. Raabe, N. Zaafarani, S. Zaefferer, Investigation of the indentation size effect through the measurement of the GND beneath small indents of different depths using EBSD tomography, Acta Mater. 57

(2008) 559–569.

[14] M.R. Begley, J.W. Hutchinson, The mechanics of size-dependent indentation, J. Mech. Phys. Solids. 46 (1998) 2049–2068.

[15] T.J. Ruggles, D.T. Fullwood, Estimations of bulk geometrically necessary dislocation density using high resolution EBSD, Ultramicroscopy. 133 (2013) 8–15.

[16] W. He, W. Ma, W. Pantleon, Microstructure of individual grains in cold-rolled aluminium from orientation inhomogeneities resolved by electron backscattering diffraction, Mater. Sci. Eng. A. 494 (2008) 21–27.

[17] W. Pantleon, Resolving the geometrically necessary dislocation content by conventional electron backscattering diffraction, Scr. Mater. 58 (2008) 994–997. doi:10.1016/j.scriptamat.2008.01.050.

[18] B.S. El-Dasher, B.L. Adams, A.D. Rollet, Viewpoint: experimental recovery of geometrically necessary dislocation density in polycrystals, Scr. Mater. 48 (2003) 141–145.

[19] D.P. Field, P.B. Tribedi, S.I. Wright, M. Kumar, Analysis of local orientation gradients in deformed single crystals, Ultramicroscopy. 103 (2005) 33–39.

[20] A.J. Wilkinson, G. Meaden, D.J. Dingley, High resolution mapping of strains and rotations using electron backscatter diffraction, Mater. Sci. Technol. 22 (2006) 1271–1278.

[21] A.J. Wilkinson, G. Meaden, D.J. Dingley, Strain mapping using Electron Backscatter Diffraction, in: Electron Backscatter Diffr. Mater. Sci., Springer, Boston, MA, 2009: pp. 231–249.

[22] T.B. Britton, A.J. Wilkinson, High resolution electron backscatter diffraction measurements of elastic strain variations in the presence of larger lattice rotations, Ultramicroscopy. 114 (2012) 82–95.

[23] T. Zhang, D.M. Collins, F.P.E. Dunne, B.A. Shollock, Crystal plasticity and high-resolution electron backscatter diffraction analysis of full-field polycrystal Ni superalloy strains and rotations under thermal loading, Acta Mater. 80 (2014) 25–38.

[24] H. Abdolvand, A.J. Wilkinson, Assessment of residual stress fields at

deformation twin tips and the surrounding environments, Acta Mater. 105 (2016) 219–231.

[25] Y. Guo, T.B. Britton, A.J. Wilkinson, Slip band-grain boundary interactions in commercially-purity titanium, Acta Mater. 76 (2014) 1–12.

[26] J. Jiang, T.B. Britton, A.J. Wilkinson, Measurement of geometrically necessary dislocation density with high resolution electron backscatter diffraction: Effects of detector binning and step size, Ultramicroscopy. 125 (2013) 1–9.

[27] N. Zaafarani, D. Raabe, R.N. Singh, F. Roters, S. Zaefferer, Three-dimensional investigation of the texture and microstructure below a nanoindent in a Cu single crystal using 3D EBSD and crystal plasticity finite element simulations, Acta Mater. 54 (2006) 1863–1876. doi:10.1016/J.ACTAMAT.2005.12.014.

[28] B.C. Larson, W. Yang, G.E. Ice, J.D. Budai, J.Z. Tischler, Three-dimensional X-ray structural microscopy with submicrometer resolution, Nature. 415 (2002) 887–890.

[29] H. Simons, A. King, W. Ludwig, C. Detlefs, W. Pantleon, S. Schmidt, F. Stöhr, I. Snigireva, Snigirev A., H.F. Poulsen, Dark-field X-ray microscopy for multiscale structural characterization, Nat. Commucation. 6 (2015).

[30] J.C.M. Li, Petch relation and grain boundary sources, Trans. Metall. Soc. AIME. 227 (1963) 239–247.

[31] J.P. Hirth, R.W. Balluffi, On grain boundary dislocations and ledges, Acta Metall. 21 (1973) 929–942.

[32] Z. Zhang, D.S. Balint, D.P.E. Dunne, Investigation of slip transfer across HCP grain boundaries with application to cold dwell fatigue, Acta Mater. 127 (2017) 43–53.

[33] M.D. Sangid, T. Ezaz, H. Sehitoglu, I.M. Robertson, Energy of slip transmission and nucleation at grain boundaries, Acta Mater. 59 (2011) 283–296.

[34] L. Wang, P. Eisenlohr, Y. Yang, T.R. Bieler, M.A. Crimp, Nucleation of paired twins at grain boundaries in titanium, Scr. Mater. 63 (2010) 827–830.

[35] Y. Guo, D.M. Collins, E. Tarleton, F. Hofmann, J. Tischler, W. Liu, R. Xu, A.J. Wilkinson, T.B. Britton, Measurement of stress fields near a grain boundary: Exploring blocked arrays of dislocations in 3D, Acta Mater. 96 (2015) 229–236.

[36] G.E. Ice, J.W.L. Pang, Tutorial on X-ray micro-Laue diffraction, Mater. Characterisation. 60 (2009) 1191–1201.

[37] W. Liu, P. Zschack, J. Tischler, G. Ice, B. Larson, X-ray laue diffraction microscopy in 3d at advanced photon source, in: AIP Conf. Proc. 1365, 2011.

[38] A.J. Wilkinson, G. Meaden, D.J. Dingley, High-resolution elastic strain measurement from electron backscatter diffraction patterns: new level of sensitivity, Ultramicroscopy. 106 (2006) 307–313.

[39] T.B. Britton, A.J. Wilkinson, Measurement of residual elastic strain and lattice rotations with high resolution electron backscatter diffraction, Ultramicroscopy. 111 (2011) 1395–1404. http://www.sciencedirect.com/science/article/pii/S0304399111001665.

[40] S. Das, F. Hofmann, E. Tarleton, Consistent determination of geometrically necessary dislocation density from simulations and experiments, Int. J. Plast. 109 (2018) 18–42.

[41] T.B. Britton, A.J. Wilkinson, Stress fields and geometrically necessary dislocation density distributions near the head of a blocked slip band, Acta Mater. 60 (2012) 5773–5782. doi:10.1016/J.ACTAMAT.2012.07.004.

[42] A.J. Wilkinson, D. Randman, Determination of elastic strain fields and geometrically necessary dislocation distributions near nanoindentation, Philos. Mag. 90 (2010) 1159–1177.

[43] V.A. Lubarda, J.A. Blume, A. Needleman, An analysis of equilibrium dislocation distributions, Acta Metall. Mater. 41 (1993) 625–642.

[44] A. Vilalta-Clemente, G. Naresh-Kumar, M. Nouf-Allehiani, P. Gamarra, M.A. di Forte-Poisson, C. Trager-Cowan, A.J. Wilkinson, Cross-correlation based high resolution electron backscatter diffraction and electron channelling contrast imaging for strain mapping and dislocation distributions in InAlN thin films, Acta Mater. 125 (2017) 125–135.

[45] R. Li, Q. Xie, Y. Wang, W. Liu, M. Wang, G. Wu, X. Li, M. Zhang, Z. Lu, C. Geng, T. Zhu, Unravelling submicron-scale mechanical heterogeneity by three-dimensional X-ray microdiffraction, Proc. Natl. Acad. Sci. 115 (2018) 483–488.

[46] F.C. Frank, W.T. Read, Multiplication Processes for Slow Moving Dislocations, Phys. Rev. 79 (1950) 722–723.

[47] H. Li, D.E. Mason, T.R. Bieler, C.J. Boehlert, M.A. Crimp, Methodology for estimating the critical resolved shear stress ratios of alpha-phase Ti using EBSD-based trace analysis, Acta Mater. 61 (2013) 7555–7567.

[48] J.P. Hirth, The influence of grain boundary on mechanical properties, Metall. Trans. 3 (1972).

[49] C.T. Young, J.L. Lytton, Computer Generation and Identification of Kikuchi Projections, J. Appl. Physics2. 43 (2003) 1408–1417.

[50] J.J. Jonas, S. Mu, T. Al-Samman, G. Gottstein, L. Jiang, Ė. Martin, The role of strain accommodation during the variant selection of primary twins in magnesium, Acta Mater. 59 (2011) 2046–2056. doi:10.1016/J.ACTAMAT.2010.12.005.

[51] T.R. Bieler, P. Eisenlohr, F. Roters, D. Kumar, D.E. Mason, M.A. Crimp, D. Raabe, The role of heterogeneous deformation on damage nucleation at grain boundaries in single phase metals, Int. J. Plast. 25 (2009) 1655–1683.

[52] L. Wang, Y. Yang, P. Eisenlohr, T.R. Bieler, M.A. Crimp, D.E. Mason, Twin nucleation by slip transfer across grain boundaries in single phase metals, Met. Mat Trans A2. 41 (2010).

[53] L. Wang, Z. Zheng, H. Phukan, P. Kenesei, J.S. Park, J. Lind, R.M. Suter, T.R. Bieler, Direct measurement of critical resolved shear stress of prismatic and basal slip in polycrystalline Ti using high energy X-ray diffraction microscopy, Acta Mater. 132 (2017) 598–610.

[54] L.E. Murr, Some observations of grain boundary ledges and ledges as dislocation sources in metals and alloys, Met. Trans. A. 6A (1975) 505–513.

[55] T.C. Lee, I.M. Robertson, H.K. Birnbaum, TEM in situ deformation study

of the interaction of lattice dislocations with grain boundaries in metals, Philos. Magzine A. 62 (1989) 131–151.

[56] J. Kacher, I.M. Robertson, Quasi-four-dimensional analysis of dislocation interactions with grain boundaries in 304 stainless steel, Acta Mater. 60 (2012) 6657–6672.

[57] J. Kacher, B.P. Eftink, B. Cui, I.M. Robertson, Dislocation interactions with grain boundaries, Curr. Opin. Solid State Mater. Sci. 18 (2014) 227–243.

[58] M.P. Dewald, W.A. Curtin, Multiscale modelling of dislocation/grain boundary interactions. II. Screw dislocations impinging on tilt boundaries in Al, Philos. Mag. 87 (2007) 4615–4641.

[59] C.W. Price, J.P. Hirth, A mechanism for the generation of screw dislocations from grain-boundary ledges, Mater. Sci. Eng. 9 (1972) 505–513.

[60] A. Serra, D.J. Bacon, R.C. Pond, Twins as barriers to basal slip in hexagonal close packed metals, Met. Mater.2 Trans. A. 33A (2002) 809–812.